\documentclass[12pt,letterpaper]{JHEP3}
\usepackage{cite}

\title{Harmonic resolution as a holographic quantum number}

\author{%
Raphael Bousso\\ 
Center for Theoretical Physics, Department of Physics\\
University of California, Berkeley, CA 94720-7300, U.S.A.\\
{\em and}\\
Lawrence Berkeley National Laboratory, Berkeley, CA 94720-8162, U.S.A.\\ 
E-mail: \email{bousso@lbl.gov}}

\abstract{%
The Bekenstein bound takes the holographic principle into the realm of
flat space, promising new insights on the relation of
non-gravitational physics to quantum gravity.  This makes it important
to obtain a precise formulation of the bound.  Conventionally, one
specifies two macroscopic quantities, mass and spatial width, which
cannot be simultaneously diagonalized.  Thus, the counting of
compatible states is not sharply defined.  The resolution of this and
other formal difficulties leads naturally to a definition in terms of
discretized light-cone quantization.  In this form, the area
difference specified in the covariant bound converts to a single
quantum number, the harmonic resolution $K$.  The Bekenstein bound
then states that the Fock space sector with $K$ units of longitudinal
momentum contains no more than $\exp(2\pi^2 K)$ independent discrete
states.  This conjecture can be tested unambiguously for a given
Lagrangian, and it appears to hold true for realistic field theories,
including models arising from string compactifications.  For large
$K$, it makes contact with more conventional but less well-defined
formulations.}

\preprint{\hepth{0310223} \\ UCB-PTH-03/26}

\begin{document}

\section{Introduction}

It was recently shown~\cite{Bou03} that the Bekenstein
bound~\cite{Bek74,Bek81} can be derived from a generalized
form~\cite{FMW} of the covariant bound on the entropy of
lightsheets~\cite{ceb1}.  This derivation becomes exact for weakly
gravitating systems in flat space.  It yields
\begin{equation}
S\leq \pi M a/\hbar,
\label{eq-gbb}
\end{equation}
where $S$ is the entropy of a matter system with energy up to $M$ and
spatial width up to $a$.

The width $a$ is the distance between any pair of parallel planes
clamping the system.  For example, if the system fits into a
rectangular box, $a$ can be taken to be its shortest side.  Hence,
(\ref{eq-gbb}) is actually stronger than Bekenstein's original
version, $S\leq \pi M d/\hbar$, which referred to the diameter $d$ of
the smallest sphere capable of enclosing the system.

Earlier derivations of the Bekenstein bound applied the generalized
second law of thermodynamics to systems that are slowly lowered into a
black hole, prompting a controversy about the role of quantum effects
and other subtleties arising in this rather nontrivial process. The
new derivation of Bekenstein's bound is immune to such difficulties as
it takes place in the benign environment of flat space and involves no
accelerations.  

Most importantly, the new derivation identifies the Bekenstein bound
as a special limit of the covariant bound~\cite{ceb1}, a conjectured
empirical pattern underlying the holographic
principle~\cite{Tho93,Sus95,ceb2}.  This limit is both intriguing and
especially simple because it applies to weakly gravitating systems.
It can be tested entirely within quantum field theory, without
inclusion of gravity.  Moreover, as we will argue in a separate
publication~\cite{BouTA}, the absence of Newton's constant in the
Bekenstein bound signifies that key aspects of quantum mechanics can
be derived from classical gravity together with the holographic
relation between information and geometry.  Hence, it will be of great
importance to obtain a completely well-defined and unambiguous
formulation of the Bekenstein bound.

Of course, our understanding of the Bekenstein bound is no worse than
that of the covariant bound.  However, for the purposes of the
covariant bound~\cite{ceb1}, the entropy $S$ can be satisfactorily
defined as the logarithm of the number of independent quantum states
compatible with assumed macroscopic conditions.  Such conditions, at
the very least, are always implicit in the specification of the area
appearing on the right hand side of the bound.  Because this area must
be large in Planck units, the bound can only be challenged by systems
with large entropy.  This is why in most situations that are of
interest for testing the bound, such as in cosmology and for
macroscopic isolated systems, thermodynamic approximations are valid
and the value of $S$ is not sensitive to subtleties (such as the
definition of ``compatible'').

By contrast, the Bekenstein bound and (by extension) the generalized
covariant bound~\cite{FMW} are most readily challenged by systems with
few quanta.  This makes them sensitive to the details of the entropy
definition.  Indeed, various authors, using inequivalent definitions,
have reached different conclusions about the validity of the
Bekenstein bound~\cite{Bek81,UnrWal82,Unw82,Pag82,Bek82,Bek83,Bek84,%
UnrWal83,SchBek89,SchBek90,Bek94,Bek94b,Bek99,%
Pag00a,Pag00b,Pag00c,Bek00b}.  Our point of view is that any concise
formulation that renders the Bekenstein bound well-defined,
nontrivial, and empirically true will capture a potentially
interesting fact about Nature.  Moreover, it may have implications in
the general context of the covariant bounds, and it may help us
sharpen their definitions as well.  Hence, we use a variety of
considerations to seek such a definition.

We have recently argued~\cite{Bou03a} that $S$ should be defined
microcanonically, as the logarithm of the number of exact eigenstates
of the Hamiltonian with energy $E\leq M$ and spatial width no greater
than $a$.  In particular, only bound states (states with discrete
quantum numbers) contribute to the entropy, since scattering states
have infinite size, and the only alternative---{\em ad hoc\/}
imposition of boundary conditions---can be shown to trigger violations
of the bound.

This definition, summarized in Sec.~\ref{sec-old}, is quite successful
heuristically.  However, it does retain one annoying ambiguity
(Sec.~\ref{sec-bad}): The spatial width of a quantum bound state is
not sharply defined.  Though wavefunctions tend to be concentrated in
finite regions, they do not normally have strictly compact support.
For example, there is a tiny but nonzero probability to detect the
electron a meter away from the proton in the ground state of hydrogen.
Of course, the width can be assigned some rough value corresponding to
the region of overwhelming support.  But this forces us to answer the
sharp question of whether or not a given state contributes to $S$ by
an inherently ambiguous decision whether the state can be considered
to have width smaller than $a$.

This problem is compounded by a practical difficulty: the Hamiltonian
methods required for the computation of bound states are often
intractable in quantum field theory.  Moreover, we show that aspects
of the formulation of Bekenstein's bound have no justification from
the point of view of its more recent derivation (which we regard as
its real origin).  Specifically, we criticize that not one but two
macroscopic parameters are specified, and that these parameters act
only to limit, but not to fix, the mass and size of allowed states.

In Sec.~\ref{sec-edlcq} we systematically develop modifications
designed to resolve these problems.  Guided by the derivation of
Bekenstein's bound from the GCEB, we construct a Fock space of states
directly on the light-sheet via light-cone quantization.  This allows
us to identify the surplus parameter in the bound as a pure gauge
choice.  Moreover, light-cone quantization famously facilitates the
use of Hamiltonian methods in quantum field theory.  Two other
problems, most notably the width ambiguity, remain.

However, in the light-cone frame, one can adopt a different gauge
which fixes the maximum width of states instead of the total momentum.
In this gauge it becomes possible to identify the light-sheet
periodically on a null circle of fixed length.  Quantization on this
compactified background is known as discretized light-cone
quantization (DLCQ).  One of its simplifying features, much exploited
in QCD calculations, is that the Fock space breaks up into distinct
sectors preserved by interactions, so that the Hamiltonian can be
diagonalized in each of them separately.  Each sector is characterized
by the number of units of momentum along the null circle, $K$.

The integer $K$ (the ``harmonic resolution'') subsumes the two
macroscopic parameters $M$ and $a$.  The Fock space contains a finite
number of bound energy eigenstates for each integer $K$.  The entropy
$S$ is defined to be the logarithm of that number, and the
Bekenstein bound takes the form
\begin{equation}
S\leq 2\pi^2 K
\end{equation}
in DLCQ.

In this form the bound is unambiguously defined and free of all of the
earlier problems we had identified.  The width of quantum states is
imposed by the compactification.  The bound manifestly contains only
one parameter, $K$, to which all contributing microstates correspond
exactly.  Because of the further simplification of the Fock space
structure, DLCQ is even better suited for finding bound states than
ordinary light-cone quantization.  Thus, all of the shortcomings we
identified are resolved.

An interesting question is whether the refined definition of entropy
developed here for flat space can be lifted back to the more general
environment in which the covariant bounds operate.  Here we hit upon a
puzzle.  Since our prescription involved compactifying a null
direction (or equivalently, demanding periodicity), it does not
naturally extend to curved space.  When the contraction of a
light-sheet cannot be neglected, its generators cannot be periodically
identified.

It is intriguing that by demanding a completely unambiguous
formulation of the Bekenstein bound, and taking seriously that entropy
bounds are tied to null surfaces, one is naturally led to the
framework of discretized light-cone quantization.  Traditionally, DLCQ
has been considered no more than a convenient trick for simplifying
numerical calculations in QCD.  More recently, it appeared in a more
substantial role in the context of the Matrix model of
M-theory~\cite{BanFis96,Sus97}.  Its independent emergence in the
context of entropy bounds suggests that DLCQ may have wider
significance.  If this were the case, then the spectra at finite
harmonic resolution may have a direct physical interpretation.

\section{Defining entropy}
\label{sec-old}

We will now discuss our starting point for the definition of entropy
in the Bekenstein bound.  In Ref.~\cite{Bou03a}, a combination of
formal and empirical arguments led us to adopt a definition in which
only bound states contribute to the entropy.  That is,
\begin{equation}
S(M,a)\equiv\log{\cal N}(M,a),
\label{eq-di}
\end{equation} 
where ${\cal N}$ is the number of independent eigenstates of the
Hamiltonian, with energy eigenvalue
\begin{equation}
E\leq M,
\label{eq-em}
\end{equation}
total three-momentum eigenvalue
\begin{equation}
{\mathbf P}=0,
\label{eq-p0}
\end{equation}
and with spatial support over a region of width no larger than $a$.
The bound takes the form
\begin{equation}
S(M,a)\leq\pi Ma/\hbar.
\label{eq-z4}
\end{equation}
We now summarize the arguments for this formulation.

The restriction to exact energy eigenstates is motivated not only by
the conceptual clarity of the microcanonical ensemble~\cite{BekSch90}.
The bound explicitly contains the mass (and not, for example, a
temperature) on the right hand side.  Thus, energy is a natural
macroscopic parameter to which microstates must conform, via
Eq.~(\ref{eq-em}).  Moreover, in the derivation of the Bekenstein
bound from the GCEB, the mass enters explicitly as the source of
focussing of light rays; no other thermodynamic quantities appear.
There are also empirical reasons: alternative definitions (involving,
for example, ensembles at fixed temperature~\cite{Deu82} or mixed
states constructed from states other than
energy-eigenstates~\cite{Pag00c}) were found to lead to violations of
the bound.

Obviously, the bound is nontrivial only for states with finite width
$a$.  Yet, we expect energy eigenstates to be spread over all of
space.  Indeed, for states which are also eigenstates of the total
spatial momentum, the overall phase factor corresponding to the
total momentum signifies a complete delocalization of the center of
mass.  This conundrum can be resolved by integrating over all spatial
momenta.  In practice, it is simpler to continue to work with
eigenstates of the full four-momentum, but to factor out and ignore
the center of mass coordinates.  We demand only that the wavefunction
have finite spreading in the position space relative to the center of
mass.  

In free field theory, however, the constituents of multi-particle
states are not bound, but are delocalized relative to each other.
Therefore, the bound is essentially trivial in free field theory:
multi-particle states have infinite spatial width and do not
contribute to the entropy.  One way of enforcing finite width would be
to impose rigid boundary conditions by fiat.  This type of
prescription leads to apparent violations of the bound~\cite{Bou03a}.
In fact it is physically incomplete, because the material enforcing
the assumed boundary conditions (for example, a capacitor with enough
charge carriers~\cite{Bek00b}) is not included in the mass and width.

Therefore the finite width requirement can be satisfied only if
interactions are properly included from the start.  Real matter
systems localize themselves by the mutual interactions of constituent
particles.  In situations where the bound has nontrivial content, this
implies that ${\cal N}$ counts energy eigenstates with finite spatial
width.  In other words, the only contribution to ${\cal N}(M,a)$ comes
from bound states, which have no continuous quantum numbers.  This
statement can be thought of as a precise version of Bekenstein's
requirement~\cite{Bek82,Bek00b} that only ``complete systems'' be
considered.

In Ref.~\cite{Bou03a} this conclusion was supported by an empirical
analysis.  We began with a free scalar and imposed boundary conditions
by fiat.  Then we estimated the mass of the materials necessary for
enforcing them.  Lower bounds on the mass of these additional
components were obtained in two different ways, using different
necessary conditions for localization.  We found that only one such
condition---the need for interactions so that particles can
bind---gives rise to extra energy sufficient to uphold the bound in
each of a diverse set of problematic examples~\cite{Bou03a}.  The
study of incomplete systems thus informs us that interactions should
be key to the definition of a complete system.

Each bound state gives rise to a continuous three-parameter set of
energy eigenstates related by boosts.  Since these states all
represent the same physical state in different coordinate systems, we
should not count them separately, but mod out by overall boosts.
Usually this is done implicitly by picking a Lorentz frame and
declaring it to be a rest frame of the ``system''.  The condition
(\ref{eq-p0}) formalizes this requirement by requiring that the
spatial components of the total four-momentum of each allowed state
must vanish.

\section{Problems of the present formulation}
\label{sec-bad}

The form (\ref{eq-di}), (\ref{eq-z4}) is an improvement over less
precise (or obviously incorrect) statements of the Bekenstein bound, but
it is still not satisfactory.  We will now identify some of its
shortcomings.  We list four problems: one ambiguity, one practical
difficulty, and two formal shortcomings.

\subsection{Width ambiguity}
\label{sec-wa}

This is the most pernicious problem because it renders the entropy $S$
manifestly ambiguous and appears to invite violations of the bound.

Energy eigenvalues are precisely defined, but the spatial width of a
bound energy eigenstate is an ambiguous concept.  In order to define a
width at all, one has to ignore the overall phase factor corresponding
to the complete delocalization of the center of mass.  One can ask,
however, about the spreading of the wave function in the remaining
position space relative to the center of mass.  As we discussed in
Sec.~\ref{sec-old}, this spreading is infinite for scattering states,
but finite for bound states.  However, wave functions of bound states
do not normally have strictly compact support in this position space;
generically, one expects at least exponential tails outside any finite
region.  How are we to define the width of such a state precisely?

One possibility is to call a state localized to a spatial region
${\cal V}$ of width $a$ if it is unlikely to find any of its
constituents outside of ${\cal V}$, i.e., if the normalized
wavefunction obeys a condition of the form $1-\int_{\cal V}
|\Psi|^2<\eta$.  But this introduces an arbitrary parameter $\eta\ll
1$, to which the integer ${\cal N}(M,a)$ is necessarily somewhat
sensitive.  A second possibility, which we will also dismiss shortly,
is to modify the Fock space construction by considering theories on
flat backgrounds in which one spatial direction is compactified on a
circle of length $a$.

The problem is particularly serious for single particle states.
Multiparticle bound states can be expanded into superpositions of
product states.  The corresponding position space functions yield a
spatial width relative to the center of mass.  The center of mass
itself is always completely delocalized for momentum eigenstates, and
the corresponding overall phase factor must be ignored to get a finite
answer.  But by this definition, single particle states of free fields
would be assigned zero spatial size, leading to obvious violations of
the bound.

\subsection{Inadequacy of Hamiltonian methods}
\label{sec-hm}

Bound states are exceedingly difficult to find exactly in quantum
field theory.  In strongly coupled theories even the vacuum is highly
nontrivial and differs significantly from the Fock space vacuum of
the free theory.  For this reason, Hamiltonian dynamics is usually
abandoned in favor of a Lagrangian formulation that lends itself to
the computation of scattering amplitudes, but not of bound
states.---This does not necessarily signal a fundamental problem, but
it does appear to render the verification of the bound intractible
precisely for the theories in which it is most interesting.

\subsection{Extra macroscopic parameters}
\label{sec-emp}

This and the following objection are related to the derivation of
Bekenstein's bound from the GCEB~\cite{Bou03a}: We will show that the
statement of the Bekenstein bound in Sec.~\ref{sec-old} is
inconsistent with its covariant origin.

The entropy $S$ in the GCEB is associated with matter systems whose
energy focuses the cross-sectional area of certain light-rays by
$\Delta A=A-A'$~\cite{Bou03a}. Hence, $\Delta A$ is the natural
``macroscopic parameter'' held fixed while counting compatible states.
The derivation of the Bekenstein bound from the GCEB converts this
area difference into the product $Ma$.  This suggests that the entropy
in the Bekenstein bound should not be obtained by specifying mass and
width separately.  Only their product, $Ma$, should be held fixed as a
single macroscopic parameter, because only this product matters as far
as the amount of focussing is concerned.

To emphasize this, let us define a new dimensionless variable:
\begin{equation}
K = \frac{Ma}{2\pi\hbar},
\label{eq-k}
\end{equation}
where a factor of $2\pi$ has been inserted for later convenience.  We
define ${\cal N}(K)$ as the number of bound states with vanishing
total momentum, whose rest mass times spatial width does not exceed
$2\pi K\hbar$.  With $S(K)\equiv \log {\cal N}(K)$, the bound takes
the form
\begin{equation}
S(K)\leq 2\pi^2 K.
\label{eq-slk}
\end{equation}

Technically, this reformulation remedies our objection.  However,
Eq.~(\ref{eq-slk}) appears to lead to a messy picture, in which states
of hugely different energy ranges and spatial sizes all contribute to
the entropy for given $K$.  In particular, Eq.~(\ref{eq-slk}) rules
out the possibility of resolving the width ambiguity
(Sec.~\ref{sec-wa}) by formally compactifying on a spatial circle of
fixed length.

\subsection{Excess parameter range}
\label{sec-epr}

We have defined ${\cal N}=e^S$ as the number of states with energy and
width {\em up to} $M$ and $a$ [or with a product of energy and width
up to $K$, in the modification (\ref{eq-slk})].  However, the
derivation of the Bekenstein bound from the GCEB~\cite{Bou03a} does
not actually support the inclusion of states with less energy or
smaller size.  Whether two surfaces, or their areas $A$ and $A'$, or
only the area difference $A-A'\sim Ma\sim K$ is held fixed: in either
case, only those states should be admitted whose energy and width
correspond precisely to $K$.  But this would render the bound trivial:
except for accidental exact degeneracies, the number of states
corresponding precisely to the specified parameters would be either
zero or one.  Moreover, such a formulation would exacerbate the
earlier problem of width ambiguity.

\section{DLCQ as a precise definition of entropy}
\label{sec-edlcq}

\subsection{Assessment}
\label{sec-assess}

Two of the problems we have listed concern the fact that parts of our
definition of entropy are hard to justify from the point of view of
the GCEB.  As we turn to remedy the situation and reformulate the
Bekenstein bound, it is therefore appropriate to look to its covariant
heritage for clues.  Indeed, there is a crucial aspect of the
covariant bounds that the form (\ref{eq-di}), (\ref{eq-z4}) of
Bekenstein's bound fails to capture: The GCEB refers to quantum states
on a portion of a light-sheet~\cite{ceb1,FMW}.  That is, it applies to
a hypersurface with two spatial and one null dimension, as opposed to
a spatial volume.

Because of the restriction to energy eigenstates, the time at which
the bound is evaluated is irrelevant, but this does not mean that the
proper definition of the entropy is equally transparent in all frames.
In Sec.~\ref{sec-old}, it was implicit that the Fock space is
constructed by equal-time quantization of the field theory in the
usual manner; then the energy eigenstates conforming to the specified
macroscopic parameters are counted.  But why artificially introduce an
arbitrary time coordinate, when the light-sheet $L$ already picks out
a (null) slicing of spacetime?

It is far more natural to regard $L$ itself as a time slice, to
construct a Fock space of states on it, and to count the number of
bound states directly on the light-sheet.  The derivation of
Bekenstein's bound becomes exact in the limit $G\rightarrow 0$, i.e.,
when all curvature radii induced by matter are much larger than the
matter system itself~\cite{Bou03a}.  In this limit, $L$ does not
contract and constitutes a {\em front}~\cite{Dir49}: a null hyperplane
in Minkowski space, given for example by $t+x=$ const.  The
construction of a Fock space on this hypersurface is known as
front-form quantization (and, less appropriately but more frequently,
as ``light-cone quantization'' or quantization in the infinite
momentum frame)~\cite{FubFur65,Wei66,Sus68,BarHal68}.  We will briefly
review the formalism; then we will show how it addresses the problems
we have identified.

\subsection{Light-cone quantization}
\label{sec-lcq}

With the coordinate change
\begin{equation}
x^+ = {t+x\over\sqrt{2}},~~~x^- = {t-x\over\sqrt{2}},
\end{equation}
the metric of Minkowski space is
\begin{equation}
ds^2 = 2dx^+dx^- - (x^\perp)^2,
\label{eq-lcm}
\end{equation}
where $x^\perp$ stands for the transverse coordinates, $y$ and $z$.
The total four-momentum has components
\begin{equation}
P^+ = P_- = {E+P^x\over\sqrt{2}},~~~P^- = P_+ = {E-P^x\over\sqrt{2}},
\label{eq-pp}
\end{equation}
and transverse components $P^\perp=(P^y,P^z)$.

In light-cone quantization, $x^+$ plays the role of time, whereas the
longitudinal coordinate $x^-$ replaces the third spatial variable.
The momentum compontent $P_+$ plays the role of a Hamiltonian; $P_-$
is called the longitudinal momentum.  Both quantities are
positive definite.

One-particle states are created by acting on the vacuum with operators
$a^\dagger_{k_- k_\perp}$ corresponding to modes
\begin{equation}
u_{k_- k_\perp} \sim\exp(ik_+ x^+ + ik_- x^- + i k_\perp x_\perp).
\end{equation}
(We suppress extra indices distinguishing different fields and
additional scalar, vector, or matrix factors for normalization and
components.)  After integrating out all zero-modes\footnote{This may
generate additional potential terms which capture nontrivial aspects
of the structure of the vacuum such as symmetry
breaking~\cite{BigSus97}.} the one-particle longitudinal momentum,
$k_-$, is strictly positive.  The one-particle light-cone energy is
given by
\begin{equation}
k_+ = \frac{m^2+k_\perp^2}{2 k_-}.
\label{eq-kp}
\end{equation}

Because of the positivity of $k_-$, and because $P_-$ is conserved,
all interaction terms contain at least one annihilation operator.
There are no terms like
$a^\dagger_{k_{-,1}}a^\dagger_{k_{-,2}}a^\dagger_{k_{-,3}}$.  Hence,
there are no radiative corrections to the vacuum, and the Fock space
can be constructed just as in the free theory.  (This constitutes one
of the chief advantages of light-cone quantization.)

As usual, the Fock space consists of products of one-particle states
obtained by acting several times with creation operators.  The free
part of the Hamiltonian takes the form
\begin{equation}
H_{\rm free}= \int_0^\infty dk_- \int d^2k_\perp~k_+~
a^\dagger_{k_-k_\perp}a_{k_-k_\perp} .
\label{eq-h}
\end{equation}
Bound states are eigenstates of the full Hamiltonian with no
continuous quantum numbers.  Bound states can be represented by
wavefunctions that describe their decomposition into the Fock space
states.  Thus, light-cone quantization permits a Lorentz-invariant
constituent interpretation of bound states even in strongly coupled
theories such as QCD~\cite{BroPau97}.

\subsection{The Bekenstein bound in front form}
\label{sec-blcq}

Let us now formulate Bekenstein's bound in the light-cone frame.  We
go back to its derivation from the covariant bound, from which the
Bekenstein most directly emerges in the covariant form~\cite{Bou03}
\begin{equation}
S\leq\pi (P_a k^a) \Delta\alpha/\hbar.
\label{eq-pka}
\end{equation}
Here $\alpha$ is an affine parameter along the generators of the
light-sheet $L$, and $\Delta\alpha$ is the length of the partial
light-sheet occupied by the matter system in question, i.e., the
``affine width'' of the system as seen by a set of parallel
light-rays.  $k^a = dx^a/d\alpha$ is the future-directed null vector
tangent to the light-sheet, and $P_a$ is the total
four-momentum~\cite{Bou03} of the matter system.\footnote{Here $P^a$
is defined so that its components correspond to the energy and the
physical momentum components, e.g., $P^x>0$ for a particle moving in
the positive $x$-direction.  We choose the metric signature $(+---)$
used in most of the field theory literature on light-cone
quantization.  By contrast, in Ref.~\cite{Bou03} the usual $(-+++)$
convention was used, and $-P^a$ stood for the physical energy-momentum
four-vector, so Eq.~(\ref{eq-pka}) took the same form.}

In Ref.~\cite{Bou03} this expression was further simplified by
specializing to an arbitrary rest frame.  Then the spatial momentum
components vanish, the affine width becomes ordinary spatial width,
and one obtains Eq.~(\ref{eq-gbb}).  We will now express
Eq.~(\ref{eq-pka}) in light-cone coordinates instead.

We take the light-sheet $L$ to be the null hypersurface $x^+=$
const.\footnote{This differs from Ref.~\cite{Bou03}, where the
light-sheet was the hypersurface $t-x=0$.  The change is made to
conform to the usual choice of surfaces of constant time in the
light-cone quantization literature.}  In the form (\ref{eq-pka}),
the Bekenstein bound is invariant under rescaling of the affine
parameter.  We choose $\alpha=x^-$, so that the affine width is
\begin{equation}
\Delta\alpha = \Delta x^-.
\end{equation}
Then the tangent vector $k^a=dx^a/d\alpha$ has components $(0,1,0,0)$
in the metric (\ref{eq-lcm}).  The expression $(P_a k^a)$ is thus
simply the longitudinal momentum $P_-$, and Eq.~(\ref{eq-pka}) takes
the form
\begin{equation}
S\leq \pi P_- \Delta x^-/\hbar.
\label{eq-blcq}
\end{equation}

Note that the light-cone energy, $P_+$, does not appear in the bound.
It is also independent of the value of the transverse momenta,
$P_\perp$.  $P_-$ and $\Delta x^-$ aquire opposite factors under
boosts, so that the product $P_- \Delta x^-$ remains invariant.
Indeed, boosts can be interpreted simply as a rescaling of the affine
parameter.  In this sense, manifest Poincar\'e invariance, though
spoiled when specializing to a spatial frame, is nearly retained by
the front form expression (\ref{eq-blcq}).  That is, Poincar\'e
transformations have no effect on Eq.~(\ref{eq-blcq}) except for
rescalings of the affine parameter.

So far, we have only expressed the bound in a new coordinate system.
Next, we turn to the question of defining $S$ in the light-cone frame.
Here we reap some benefits that allow us to address two of the four
shortcomings listed in Sec.~\ref{sec-bad}.

The direct analogue of the prescription (\ref{eq-di}) would be to
specify two macroscopic parameters, $P_-$ and $\Delta x^-$, and to
define the entropy by
\begin{equation}
S =\log {\cal N}_{LCQ}(P_-,\Delta x^-),
\label{eq-slcq}
\end{equation}
where ${\cal N}_{LCQ}(P_-,\Delta x^-)$ is the number of eigenstates of
the total four-momentum, whose longitudinal momentum and affine
width do not exceed the specified parameters.

We also require an analogue of the gauge condition, Eq.~(\ref{eq-p0}),
to ensure that states related by overall boosts are counted only once.
This condition can be adapted to the light-cone frame by fixing those
components of the four-momentum which are canonically treated as
spatial, namely $P_-$ and $P_\perp$.  The transverse momenta can be
set to zero as before:
\begin{equation}
P_\perp=0,
\label{eq-perp0}
\end{equation}
which projects out states related by transverse boosts.  However, one
of the peculiarities of the light-cone frame is that the longitudinal
momentum is strictly positive for massive states.  It cannot be
gauge-fixed to zero by boosting.  In order to mod out by longitudinal
boosts, $P_-$ must instead be set to an arbitrary positive constant:
\begin{equation}
P_- = {\rm const}.
\label{eq-pmg}
\end{equation}

This exposes the ``macroscopic parameter'' $P_-$ specified in the
definition (\ref{eq-slcq}) of the entropy as a gauge choice.  Only the
width $\Delta x^-$ is a physical parameter.  Thus, when the bound is
formulated in the light-cone frame, the existence of only one
macroscopic parameter is manifest, and the objection in
Sec.~\ref{sec-emp} is resolved.

In Sec.~\ref{sec-hm} we objected that Hamiltonian methods, which are
crucial to the identification of bound states and thus to our
definition of entropy, are impractical and hardly used in quantum
field theory.  But in fact, light-cone quantization facilitates the
use of Hamiltonians considerably.  For example, the light-cone
Hamiltonian $P_+$, unlike the energy $E$, can be evaluated from the
other four-momentum components without use of a square root; see
Eqs.~(\ref{eq-kp}) and (\ref{eq-h}).  Moreover, the ground state of
the free theory is also a ground state of the interacting Hamiltonian.
For these and other reasons, light-cone quantization has emerged as a
leading tool for finding the spectrum and wavefunctions of bound
states in QCD and other interacting theories~\cite{BroPau97}.
Although we were guided to the front form by a different consideration
(the covariant pedigree of Bekenstein's bound), we thus find that
light-cone quantization is custom-designed for the task of defining
the relevant entropy.

\subsection{Resolving the width ambiguity by compactification}
\label{sec-comp}

Having succeeded in resolving two of the four problems identified in
Sec.~\ref{sec-bad}, we now turn to the two remaining difficulties---in
particular, the dreaded width ambiguity.

Let us rewrite Eq.~(\ref{eq-blcq}) in the manifestly Lorentz-invariant
form:
\begin{equation}
S(K)\leq 2\pi^2 K,
\end{equation}
where $K$ is an arbitrary non-negative number specified as a
macroscopic parameter.  In the light-cone frame, $K$ is given by
\begin{equation}
K=\frac{P_-\Delta x^-}{2\pi \hbar}.
\end{equation}
Once $P_-$ is gauge-fixed to a constant, specification of the
parameter $K$ is equivalent to specification of $\Delta x^-$.

Its manifest boost invariance allows us to think of the front form of
Bekenstein's bound in two ways.  In both versions, $K$ is the single
macroscopic parameter.  Until now we have chosen to gauge-fix $P_-$
and obtain a bound for every positive $K$, concerning the entropy of
states whose maximal width depends on $K$ as
\begin{equation}
\Delta x^- = 2\pi K\hbar/P_-.
\label{eq-xk}
\end{equation}
An alternative, equivalent option is to gauge-fix $\Delta x^-$ but
leave $P_-$ to be determined by
\begin{equation}
P_- = 2\pi K\hbar/\Delta x^- .
\end{equation}
This also yields a bound for every positive $K$, concerning the
entropy of states of fixed width but $K$-dependent maximal
longitudinal momentum.

Both pictures yield the same number of states, because every physical
state allowed for a given value of $K$ is mapped to a boosted version
of itself when the picture is changed.  But the second picture, in
which the width is gauge-fixed, serves as a point of departure for a
new formulation of the Bekenstein bound which circumvents the
ambiguity of the width of a quantum state.

We may now directly enforce a kind of width limit on quantum states
simply by compactifying the $x^-$ direction on a light-like circle of
affine length $\Delta x^-$.  This contrasts with the rest frame, in
which no such unique compactification is possible, because the spatial
width $a$ is still variable even after specifying $Ma$ and
gauge-fixing the three-momentum to zero.  Because $\Delta x^-$ can be
gauge-fixed, and can be fixed to the same value independently of $K$,
we can consistently compactify on a fixed null circle.

Note that a prescription that involves compactification is a genuine
modification of the bound.\footnote{It is thus a more radical step
than merely going to the light-cone frame, which is merely better
adapted but physically equivalent to ordinary Lorentz frames.}  It
changes the spectrum, especially at small values of $K$.  The finite
size of the longitudinal direction means that the distinction between
bound states and scattering states can only be based on the bahavior
in the transverse directions.  If this prescription is the correct
formulation of the Bekenstein bound, then the application of the bound
to real systems will require choosing $K$ so large that the effects of
compactification are negligible.  In any case, the ambiguity of
defining the spatial width of a quantum states forces a compactified
formulation upon us.

This presents us with the task of constructing a Fock space of states
on a light front with periodic boundary conditions.  Fortunately, this
formalism is well understood; indeed, discretized light-cone
quantization~\cite{PauBro85a,PauBro85b} is one of the chief tools for
calculating bound states in QCD~\cite{BroPau97}.  Let us briefly
review the key elements.

\subsection{Discretized light-cone quantization}
\label{sec-dlcq}

Compactification of the $x^-$ direction discretizes all longitudinal
momenta, which must be integer multiples of $2\pi\hbar/\Delta x^-$.
In particular, the parameter
\begin{equation}
K=\frac{P_-\Delta x^-}{2\pi \hbar}
\label{eq-kpx}
\end{equation}
is now a non-negative integer called the harmonic resolution.
The correspondingly modified Fock space construction is called
discrete light-cone quantization (DLCQ).  One-particle states still
correspond to modes
\begin{equation}
u_{k_- k_\perp} \sim\exp(ik_+ x^+ + ik_- x^- + i k_\perp x^\perp),
\end{equation}
but now their longitudinal momentum is discrete:
\begin{equation}
k_- = \frac{2\pi n\hbar}{\Delta x^-},
\end{equation}
where $n$ is a positive integer.

Because $P_-$ is conserved by interactions, the Fock space decomposes
into an infinite number of {\em inequivalent\/} sectors, one for each
nonnegative integer $K$.  Note that the one-particle states have
positive, quantized longitudinal momenta, which must add up to $2\pi
K\hbar/\Delta x^-$ in the $K$-th sector.  This makes the Fock space
sectors of DLCQ comparatively simple.  For example the $K=1$ sector
can only contain one-particle states, all of which have
$k_-=2\pi\hbar/\Delta x^-$.

\subsection{The Bekenstein bound in DLCQ form}
\label{sec-bdlcq}

Given a field theory in discretized light-cone quantization, let
${\cal N}_{\rm DLCQ}(K)$ be the number of bound states in the sector
of the Fock space with harmonic resolution $K$.  By bound states we
mean those states in the spectrum of the Hamiltonian $P_+$ which are
discrete up to overall boosts.  We define the entropy
\begin{equation}
S_{\rm DLCQ}(K)=\log {\cal N}_{\rm DLCQ}(K).
\label{eq-sdlcq}
\end{equation}
The Bekenstein bound in DLCQ form is the conjecture that
\begin{equation}
S_{\rm DLCQ}(K)\leq 2\pi^2 K.
\label{eq-dbb}
\end{equation}

For completeness we summarize the gauge conditions again.  Previously
they corresponded to fixing the total momentum, as in
Eq.~(\ref{eq-p0}), or Eqs.~(\ref{eq-perp0}) and (\ref{eq-pmg}).  In
the DLCQ formulation, we still must set the transverse momenum
components to a fixed value; for example,
\begin{equation}
P_\perp = 0.
\end{equation}
We no longer gauge-fix $P_-$; that is replaced by picking an arbitrary
but fixed compactification length $\Delta x^-$.  Note that the
spectrum depends trivially on $\Delta x^-$, and the entropy
(\ref{eq-sdlcq}) in the sector $K$ does not depend on $\Delta x^-$ at
all.

Let us summarize how the problems listed in Sec.~\ref{sec-bad} have
been resolved by formulating the Bekenstein bound in DLCQ.  The
problem of defining the width of quantum states (Sec.~\ref{sec-wa}) is
circumvented, because width enters only implicitly through the fixed
compactification scale, to which all states conform by construction.
The entropy $S$ is defined unambiguously by the specification of only
a single parameter, $K$, which corresponds to the area difference in
the GCEB, as demanded in Sec.~\ref{sec-emp}.  All states contributing
to $S$ correspond {\em precisely\/} to the sector with $K$ units of
longitudinal momentum, and not to a range (as was criticized in
Sec.~\ref{sec-epr}).  The light-cone frame is ideal for the use of
Hamiltonian methods and computation of bound states, and discrete
light-cone quantization facilitates this task further~\cite{BroPau97}.

\section{Discussion}

We have achieved our goal of obtaining a precise formulation of the
Bekenstein bound which also satisfies several formal constraints
related to its origin from bounds on light-sheets.  We were motivated
by the expectation that Bekenstein's bound captures constraints that
the holographic principle imposes on the physics of flat space---a
point of view that will be discussed in detail in a forthcoming
publication~\cite{BouTA}.  In this section, we note that the DLCQ form
of Bekenstein's bound is empirically viable.  We also point out some
implications and puzzles arising from the null compactification.

\subsection{Validity}

We expect that the Bekenstein bound in DLCQ form,
Eq.~(\ref{eq-dbb}), is valid for realistic field theories.  Many
explicit calculations of spectra in DLCQ have been carried out (see
Ref.~\cite{BroPau97} for a review), especially in the context of QCD.
In a preliminary survey, we have found no results which contradict
Eq.~(\ref{eq-dbb}).  It will be an interesting task to check the
bound systematically against existing results and to calculate more
spectra for further verification.  Because of the rapidly increasing
complexity of diagonalizing the Hamiltonian, results in the literature
pertain mostly to small values of $K$, but this is the most
interesting range in any case.  When a large number of quanta is
present, the bound tends to be easily satisfied~\cite{BouFla03}.
Violations of the bound would require a surprisingly strong growth of
the number of bound states with $K$, at low $K$.

The species problem, which appeared to be resolved by
interactions~\cite{Bou03a}, resurfaces in the DLCQ form.  One can
write down Lagrangians that populate the $K=1$ sector with an
arbitrary number $Q$ of fundamental one-particle states.  Unless the
theory is confining, the bound will thus be violated if
$Q>\exp(2\pi^2)\approx 3\times 10^8$.  We interpret this as a
prediction that Lagrangians with such a large number of fields are not
consistent with quantum gravity.  Certainly there are no indications
that such Lagrangians would be realistic.

Before we took the step of null compactification, the restriction to
bound states followed automatically from the requirement of finite
spatial width.  Now, however, it must be imposed explicitly.
Particles can scatter off to infinity in the uncompactified transverse
dimensions.  Scattering states contribute a continuous part to the
spectrum, which must be ignored when calculating the entropy.  An
interesting question, which we do not investigate here, is whether
long-lived resonances can be treated in a controlled way.  Even the
proton is probably metastable, not to speak of ordinary macroscopic
systems, to which the bound ought to apply nevertheless.  It may turn
out that such states are effectively included because they have stable
antecedents at finite $K$ where the resolution does not suffice to
describe the decay products.

Our prescription has a further restriction which, one hopes, can be
relaxed without sacrificing precision: that the transverse spatial
dimensions are noncompact.  One would like to consider not only exact
Minkowski space (with the required null identification), but also
compactifications from higher-dimensional theories.  The resulting
tower of Kaluza-Klein modes gives an infinite number of species from
the lower-dimensional point of view.  If we wish to apply the bound to
flat space with compact dimensions, it is natural to restrict to the
massless sector.  In many string compactifications, this sector can
still contain a considerable number of species ($Q\sim 10^4$), but we
are not aware of examples which exceed the bound.  Another acceptable
limit may be to consider only states which are so well localized in
the compact dimension that the situation is equivalent to
higher-dimensional flat space.  However, it is difficult to
distinguish such bound states from states which would become unstable
in the decompactification limit.

\subsection{Implications and Puzzles}

The precision gained by compactifying a null direction comes at a
price.  The spectrum in the sectors with small $K$ differs from the
true spectrum of the theory, which is strictly recovered only in the
decompactification limit $K\to\infty$.  At finite $K$, sufficiently
complex systems and fine spectral features are not resolved.

However, DLCQ does approximate physical states with $Ma/\hbar\ll K$
very well~\cite{BigSus97}.  Thus, for sufficiently large $K$, the DLCQ
form does connect with more traditional but less precise formulations
of Bekenstein's bound, in which a particular matter system with fixed
mass and size is given.

What is somewhat mysterious is whether and how our refinement of the
entropy definition lifts back to the more general light-sheets allowed
by the covariant entropy bound.  In the weak gravity limit, the
specific light-sheets chosen for the derivation of Bekenstein's bound
become a null hyperplane ($x+t=$ const).  But generically, the
cross-sectional area of light-sheets decreases.  Such light-sheets
cannot be periodically identified along the null direction.  It may be
more useful to think of DLCQ as an imposition of periodicity rather
than the physical compactification of light-rays.

The appearence of DLCQ when making Bekenstein's bound precise may
indicate that this form of quantization plays a preferred role in the
emergence of ordinary flat space physics from an underlying quantum
gravity theory (just as null hypersurfaces may have a special
significance in how general relativity arises).  If this is the case,
we will eventually discover a physical interpretation of the spectra
at finite $K$.

\acknowledgments

I would like to thank M.~Aganagic, N.~Arkani-Hamed, T.~Banks,
D.~Kabat, E.~Martinec, S.~Shenker, A.~Strominger, and L.~Susskind for
discussions.

\bibliographystyle{board}
\bibliography{all}
\end{document}